\documentclass[prb,twocolumn,showpacs]{revtex4}
\usepackage{graphicx}
\usepackage{dcolumn}
\usepackage{bm}

\begin{document}

\preprint{ATB-1}

\title{High-energy Cu spin excitations in
PrBa$_{2}$Cu$_{3}$O$_{6+x}$}

\author{A. T. Boothroyd}
\email{a.boothroyd1@physics.ox.ac.uk}
\homepage{http://xray.physics.ox.ac.uk/Boothroyd}\affiliation{
Department of Physics, Oxford University, Oxford, OX1 3PU, United
Kingdom }
\author{N. H. Andersen}\affiliation{ Ris\o National Laboratory,
DK-4000 Roskilde, Denmark }
\author{B. H. Larsen}\affiliation{ Ris\o National Laboratory,
DK-4000 Roskilde, Denmark }
\author{A. A. Zhokhov}\affiliation{
Russian Academy of Sciences, ISSP, Chernogolovka 14232, Russia }
\author{C. D. Frost}\affiliation{ ISIS Division, Rutherford Appleton Laboratory,
Didcot, Oxon., OX11 0QX, United Kingdom}
\author{D. T. Adroja}
\affiliation{ ISIS Division, Rutherford Appleton Laboratory,
Didcot, Oxon., OX11 0QX, United Kingdom}

\date{\today}

\begin{abstract}
This paper describes high-energy neutron inelastic scattering
measurements of propagating magnetic excitations in
PrBa$_{2}$Cu$_{3}$O$_{6+x}$ ($x \approx 0.2$ and $0.93$). The
measurements probe the acoustic and optic modes of the
antiferromagnetically-ordered copper--oxygen bilayers in the
energy range 50--150\,meV. The observed magnon dispersion can be
described satisfactorily in this energy range by a spin wave model
including intra- and inter-layer nearest-neighbour exchange
constants $J_{\parallel}$ and $J_{\perp}$. We find
$J_{\parallel}=127\pm 10$\,meV and $J_{\perp}=5.5\pm 0.9$\,meV.
The value of $J_{\parallel}$ is virtually the same as that found
in YBa$_{2}$Cu$_{3}$O$_{6.2}$, but $J_{\perp}$ is a factor of two
smaller. To within experimental error the values of
$J_{\parallel}$ and $J_{\perp}$ for PrBa$_{2}$Cu$_{3}$O$_{6+x}$ do
not vary with oxygen doping.
\end{abstract}

\pacs{PACS numbers: 74.72.Bk, 75.30.Ds, 75.50.Ee, 78.70.Nx}
\maketitle

\section{\label{sec:intro}Introduction}

One of the unusual characteristics of the cuprate superconductors
is the relative insensitivity of the superconducting properties to
the presence of magnetic rare-earth ions. In the {\it
R}Ba$_{2}$Cu$_{3}$O$_{6+x}$ family ({\it R} = Y or rare earth)
superconductivity occurs at temperatures as high as 95\,K, and
magnetic ordering of the {\it R} sublattice generally coexists
with superconductivity at temperatures around 2\,K or
below.\cite{Allenspach-Handbook-2001} Antiferromagnetic (AFM)
ordering of the bilayer Cu spins is observed in
non-superconducting samples with low oxygen doping levels
(typically $x<0.4$), but superconducting samples do not exhibit
any conventional form of Cu magnetic
order.\cite{Magnetic-order-YBCO}

A striking exception to this norm is the case of
PrBa$_{2}$Cu$_{3}$O$_{6+x}$ (hereafter PrBCO6+$x$), which exhibits
anomalous electrical and magnetic properties in comparison to
other {\it R}BCO
compounds.\cite{radousky,Boothroyd-JAlloysCmpds-2000}
Superconductivity is not found for any $x$ in samples of
PrBCO6+$x$ prepared by standard methods\cite{Zou-JJAP-1997}, and
the transition temperature $T_{\rm N}\approx 300$\,K for AF order
of the Cu spins depends only weakly on $x$.\cite{PrBCO-TN}
Magnetic ordering of the Pr sublattice takes place below a
temperature $T_{\rm Pr}$ varying from 11\,K ($x=0$) to 17\,K
($x=1$),\cite{PrBCO-TN,Kebede_PRB-1989,Li-PRB-1989,Wortmann-Felner-1990}
much higher than the rare earth magnetic ordering temperatures in
superconducting {\it R}BCO compounds. Unusual magnetic structures
are observed below $T_{\rm Pr}$ due to magnetic coupling between
the Cu and Pr sublattices.\cite{Boothroyd-Cu-Pr-coupling}

Many of these anomalous features of PrBCO6+$x$ remain a puzzle.
Models for the electronic structure have indicated a tendency for
Pr 4$f$ -- O 2$p$ hybridization to cause a localization of doped
holes in O 2$p_{\pi}$ orbitals, thus inhibiting
superconductivity.\cite{FR-LM-PRL} Such hybridization would also
be expected to influence the magnetic couplings in PrBCO6+$x$, and
so measurements of the exchange interactions could yield
information on the underlying electronic structure.

Over a number of years we have undertaken a systematic
investigation of the magnetic excitations of PrBCO6+$x$ by neutron
inelastic
scattering.\cite{Boothroyd-PhysicaC-1993,Lister-PRL-2001,Gardiner-ApplPhys-2002}
The aim has been to identify the important magnetic couplings and
to see how they vary with doping. Broadly speaking, the magnetic
excitations fall into two categories: (i) transitions between
levels of the Pr 4$f$ electrons split by the local crystalline
electric field
(CEF),\cite{Boothroyd-PhysicaC-1993,CEF-excitations} and (ii) Cu
spin wave excitations. There are also non-trivial effects on the
excitations due to the magnetic coupling between the Pr and Cu
sub-systems.\cite{Lister-PRL-2001,Gardiner-ApplPhys-2002}

In our most recent work we have investigated the wavevector
dependence of the magnetic excitations using a single crystal
sample of PrBCO6+$x$ prepared first in an oxygen-deficient state
($x\approx 0.20$) and later treated to produce an
`optimally-doped' oxygen content ($x\approx 0.93$). Refs.
\onlinecite{Lister-PRL-2001} and
\onlinecite{Gardiner-ApplPhys-2002} describe measurements at low
($< 10$\,meV) and intermediate (45--65\,meV) energies, and present
an analysis of the spectra in terms of a spin model for the
coupled Cu--Pr system. Here we report measurements of the Cu spin
excitations at higher energies (50--150\,meV) by time-of-flight
neutron scattering. We determine the principal intra- and
inter-layer Cu--Cu exchange interactions $J_{\parallel}$ and
$J_{\perp}$ in PrBCO6+$x$ and make a comparison with the
corresponding parameters in YBCO6+$x$. Perhaps the most
interesting new finding is that $J_{\parallel}$ and $J_{\perp}$ in
PrBCO6+$x$ are independent of oxygen doping.


\section{\label{sec:expt}Experimental details}

The experiments were performed on the same single crystal of
PrBCO6+$x$ as used for the measurements described in Refs.
\onlinecite{Lister-PRL-2001} and
\onlinecite{Gardiner-ApplPhys-2002}, and also for the phonon study
reported in Ref. \onlinecite{Gardiner-PRB-2004}. The as-grown
crystal had a mass of 2\,g, and a mosaic spread of approximately
1\,deg (full width at half maximum) measured by neutron
diffraction. The crystal was subjected to two annealing treatments
to control the oxygen content. First, the as-grown crystal was
reduced at 700$^{\circ}$C in a flow of 99.998\% argon for 100\,h
and quenched to room temperature. After the measurements on this
`under-doped' crystal had been completed the crystal was then
annealed in pure oxygen over a period of 60 days while cooling in
steps from 600$^{\circ}$C to 450$^{\circ}$C, with progressively
longer dwell times at each step as the temperature decreased.
After the second anneal the mosaic of the crystal was found to
have increased to approximately 3\,deg.

\begin{figure}
\begin{center}
\includegraphics
[width=6cm,bbllx=178,bblly=438,bburx=408,
bbury=636,angle=0,clip=]{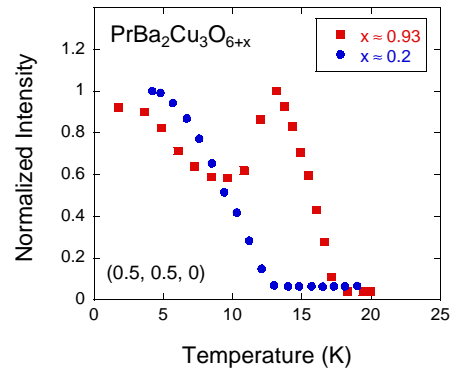} \caption{(Color online).
Temperature dependence of the $(0.5, 0.5, 0)$ magnetic Bragg peak
of PrBCO6+$x$. This peak signals magnetic order on the Pr
sublattice. The curves were recorded after the annealing
treatments described in the text, which resulted in oxygen
contents of $x \approx 0.2$ and $x \approx 0.93$. The measurements
were made by neutron diffraction on the triple-axis spectrometers
TAS6 at Ris\o\ National Laboratory and IN14 at the Institut
Laue-Langevin. The maximum intensity has been scaled to unity for
the sake of comparison. \label{fig:0} }
\end{center}
\end{figure}

According to studies of oxygen content in YBCO6+$x$ as a function
of annealing conditions (oxygen partial pressure and temperature)
the two treatments applied here yield oxygen contents of $x
\lesssim 0.2$ and $x \approx 0.93$.\cite{annealing} Although the
oxygen ordering properties of YBCO6+$x$ and PrBCO6+$x$ are
qualitatively different their oxygen equilibrium isobar coincides
at 0.21 atm up to 650$^{\circ}$C.\cite{Parfionov-Konovalov}
Therefore it is realistic to assume that this is also the case for
other oxygen partial pressures. However, as a further check of the
oxygen content we measured the magnetic ordering behaviour of the
crystal after each anneal by neutron diffraction. Studies carried
out on polycrystalline PrBCO6+$x$ have shown by various
techniques\cite{PrBCO-TN,Kebede_PRB-1989,Li-PRB-1989,Wortmann-Felner-1990}
that the Pr magnetic ordering temperature increases from $T_{\rm
Pr} = 11$\,K for vacuum-annealed samples ($x \approx 0$), to
$T_{\rm Pr} = 17$\,K for `optimally-doped' material ($x \approx
0.93$). The results for our crystal are illustrated in Fig.\
\ref{fig:0}, which shows the temperature dependence of the $(0.5,
0.5, 0)$ Bragg reflection whose appearance is known to signal the
onset of Pr magnetic
order.\cite{Boothroyd-Cu-Pr-coupling,Longmore-PRB-1996,Uma-JPCM-1998}
The Pr ordering temperatures are determined to be $T_{\rm Pr} =
13\pm 0.5$\,K after reduction, and $T_{\rm Pr} = 17.5\pm0.5$\,K
after oxygen annealing. These values are consistent with the
literature results for the expected oxygen contents $x \approx
0.2$ and $x \approx 0.93$, respectively, and we will henceforth
refer to the crystals as PrBCO6.2 and PrBCO6.93. The sharpness of
the transitions indicates a homogeneous oxygen distribution. The
$(0.5, 0.5, 0)$ intensity curve for PrBCO6.93 also shows an
anomaly at $T_2 = 12.5 \pm 0.5$. This is consistent with a spin
transition first observed by Uma {\it et al}.,\cite{Uma-JPCM-1998}
which corresponds to a change in the $c$ axis stacking sequence of
the coupled Pr--Cu magnetic
structure.\cite{Boothroyd-JAlloysCmpds-2000}

The neutron inelastic scattering measurements reported here were
performed on the MAPS (PrBCO6.2) and HET (PrBCO6.93) spectrometers
at the ISIS spallation neutron source. MAPS and HET are
time-of-flight spectrometers equipped with pixellated area
detectors situated 6\,m (MAPS) and 4\,m (HET) from the sample
position, and an incident beam divergence of approximately
0.5$^\circ$. These design features ensure very good wavevector
({\bf Q}) resolution, an essential experimental requirement in
this work because of the very steep Cu spin wave dispersion.

The crystal was aligned with the c axis approximately parallel to
the incident neutron beam direction. This arrangement means (i)
that the surfaces in {\bf Q} space corresponding to constant
neutron energy transfer are approximately parallel to the
$a^{\ast}b^{\ast}$ plane in the reciprocal lattice of the crystal,
and (ii) that the component of {\bf Q} parallel to $c^{\ast}$
varies with energy transfer. The natural way to visualize the data
is then to integrate the signal recorded in each pixel of the
detector bank over a band of energies and project the results on
to the $a^{\ast}b^{\ast}$ plane.

Measurements were performed with fixed incident neutron energies
of 80\,meV, 200\,meV and 300\,meV on MAPS, and 200\,meV on HET.
Most of the data were collected with the crystal at a temperature
of 20\,K, but some runs were repeated at 300\,K. The runs on MAPS
were of 2--3 days duration, and the runs on HET were of 5 days
duration for each of 2 settings of the crystal. These run times
are given for an average proton current of 160\,$\mu$A. In
principle, the range of incident energies employed on MAPS allowed
us to study the spin excitations up to $\sim 260$\,meV, which is
the anticipated maximum in the one-magnon spectrum (see later).
However, owing to the relatively small size of the crystal the
count rate was too low to obtain statistically meaningful data
above 150\,meV.  The intensity was converted into an absolute
cross section by comparison with the scattering from a standard
sample of vanadium.\cite{Windsor} The presented spectra are the
partial differential cross section ${\rm d}^2 \sigma/{\rm
d}\Omega{\rm d}E_{\rm f}$ per formula unit (f.u.) multiplied by
the factor $k_{\rm i}/k_{\rm f}$, where $k_{\rm i}$ and $k_{\rm
f}$ are the initial and final neutron wavevectors, and $E_{\rm f}$
is the final energy.\cite{Squires}

\section{\label{sec:inelastic}Cu spin excitation spectrum}

\begin{figure}
\begin{center}
\includegraphics
[width=5cm,bbllx=138,bblly=229,bburx=434,
bbury=698,angle=0,clip=]{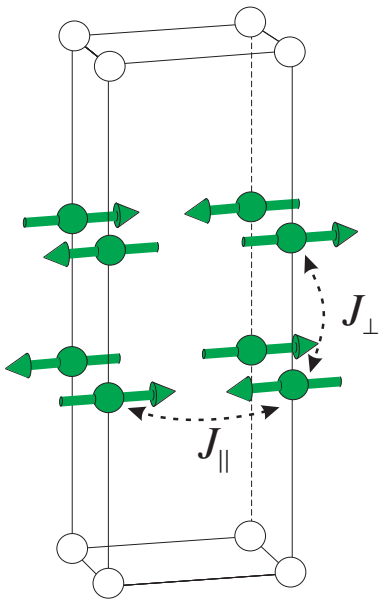} \caption{(Color online). The AF1
magnetic structure. The diagram depicts one unit cell of
PrBCO6+$x$. Only the Cu atoms are shown, and ordered magnetic
moments are carried only by the Cu atoms in the bilayer. The
constants $J_{\parallel}$ and $J_{\perp}$ are the
nearest-neighbour intra- and inter-layer exchange parameters,
respectively.\label{fig:1} }
\end{center}
\end{figure}

As mentioned above, the bilayer Cu spins in PrBCO6+$x$ order
antiferromagnetically at a temperature $T_{\rm N}$ near room
temperature and slightly dependent on $x$. At temperatures below
$T_{\rm N}$ but above $T_{\rm Pr}$ the bilayer Cu spins are
aligned antiparallel to their nearest neighbours along all three
crystallographic directions, as shown in Fig. \ref{fig:1}.  There
is no ordered moment on the Cu site in the Cu--O chains. This is
the well known AFI structure, the same as found in underdoped
YBCO6+$x$. Our interpretation of the data can therefore be guided
by the spin wave model used to describe the spin excitations in
AFI-ordered
YBCO6+$x$.\cite{Tranquada-PRB-1989,Reznik-PRB-1996,Hayden-PRB-1996}

For the high-energy spin excitations it is a good approximation to
regard the Cu spin arrangement as a square-lattice bilayer
antiferromagnet. This neglects the slight orthorhombic distortion
(if present) and the weak coupling between bilayers in adjacent
unit cells. The simplest spin Hamilatonian that describes this
system is then
\begin{equation}
H = J_{\parallel}\hspace{-2pt}\sum_{\langle ij
\rangle}\hspace{-2pt}{\bf S}_i\cdot{\bf S}_j +
J_{\perp}\hspace{-2pt}\sum_{\langle ij' \rangle}\hspace{-2pt}{\bf
S}_i\cdot{\bf S}_{j'},\label{eq:1}
\end{equation}
The summations in Eq.~(\ref{eq:1}) are over pairs of
nearest-neighbour Cu spins in the same layer (first term) and on
adjacent layers (second term). Each pair of spins is counted only
once. The constants $J_{\parallel}$ and $J_{\perp}$ are the intra-
and inter-layer exchange parameters, respectively.

The magnon spectrum derived from Eq.~(\ref{eq:1}) has two
branches, differing according to whether the spins on adjacent
layers in the bilayer rotate in the same sense (acoustic modes)
about their average direction, or in the opposite sense (optic
modes). The dispersion of these branches is given
by\cite{Tranquada-PRB-1989,Hayden-PRB-1996}
\begin{equation}
\hbar\omega({\bf Q}) = 2J_\|\{1-\gamma^2({\bf
Q})+(J_\bot/2J_\|)[1\pm\gamma({\bf Q})]\}^{1/2},\label{eq:2}
\end{equation}
where $+$ and $-$ correspond to the acoustic and optic modes,
respectively, and
\begin{equation}
\gamma({\bf Q}) =
\frac{1}{2}\{\cos(Q_xa)+\cos(Q_ya)\}.\label{eq:3}
\end{equation}
Close to the 2D AFM zone centres, e.g. ${\bf Q} =
(\frac{1}{2},\frac{1}{2})$ in units of $2\pi/a$, the acoustic
branch is linear with an initial slope proportional to
$J_{\parallel}$. The optic branch has an energy gap of
$2\surd(J_{\parallel}J_{\perp})$ at the AFM zone centres. Because
of the bilayer structure in the unit cell (Fig.\ \ref{fig:1}) the
cross section for scattering from acoustic and optic magnons
contains factors of $\sin^2(zQ_zc/2)$ and $\cos^2(zQ_zc/2)$,
respectively, where $z$ is the inter-layer spacing as a fraction
of the $c$ lattice parameter ($z=0.295$ for PrBCO6+$x$).

\section{\label{sec:results}Results}

\begin{figure}
\begin{center}
\includegraphics
[width=8.5cm,bbllx=136,bblly=528,bburx=409,
bbury=732,angle=0,clip=]{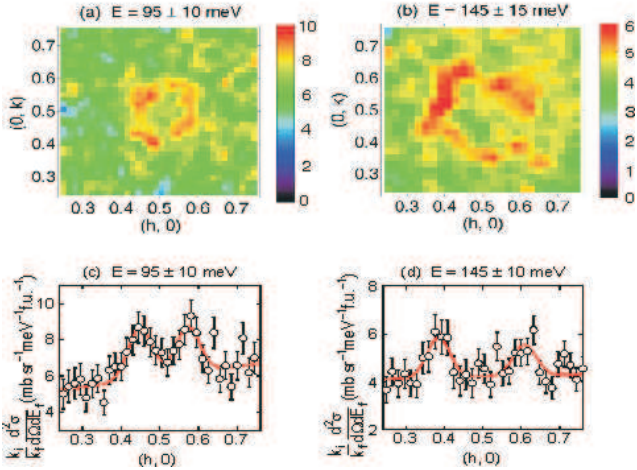} \caption{(Color online).
Neutron scattering from PrBa$_{2}$Cu$_{3}$O$_{6+x}$ ($x \approx
0.2$) measured on the MAPS spectrometer. (a) and (b) are
constant-energy slices in which the intensity has been averaged
over a band of energies from (a) 85\,meV to 105\,meV, and (b)
130\,meV to 160\,meV. (c) and (d) are cuts taken from the slices
in (a) and (b) parallel to $(h,0)$ passing through $(0.5,0.5)$.
The solid lines are the results of fitting the data to two
Gaussian functions on a sloping background. The incident neutron
energies used in the measurements were 200\,meV [(a) and (c)] and
300\,meV [(b) and (d)]. Data from the four equivalent Brillouin
zones $(\pm0.5,\pm0.5)$ have been averaged. \label{fig:2} }
\end{center}
\end{figure}
\begin{figure}
\begin{center}
\includegraphics
[width=7.5cm,bbllx=99,bblly=582,bburx=388,
bbury=727,angle=0,clip=]{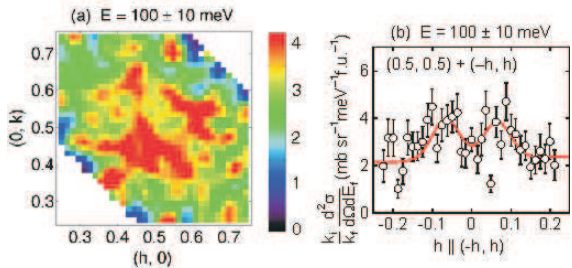} \caption{(Color online).
Neutron scattering from PrBa$_{2}$Cu$_{3}$O$_{6+x}$ ($x \approx
0.93$) measured on the HET spectrometer with an incident energy of
200\,meV. (a) is a constant-energy slice in which the intensity
has been averaged over the range 90--110\,meV, and (b) is a cut
taken parallel to the $(-h,h)$ direction passing through
$(0.5,0.5)$. The solid lines are the results of fitting the data
to two Gaussian functions on a sloping background. \label{fig:3} }
\end{center}
\end{figure}

Figures \ref{fig:2} and \ref{fig:3} present some examples of
neutron scattering data collected on the MAPS and HET
spectrometers, respectively. Figs. \ref{fig:2}(a) and (b) are maps
of the intensity from PrBCO6.2 averaged over the energy range
85--105\,meV and 130--160\,meV, respectively, projected on the
$(h,k)$ plane. For these energies we could not detect any
difference between the signal at 20\,K and at 300\,K, so data
collected at these two temperatures were averaged to improve
statistics. The images focus on the region of 2D reciprocal space
in the vicinity of the AFM zone centre $(0.5,0.5)$, and contain
intensity in excess of background distributed around $(0.5,0.5)$.
This signal, which expands away from $(0.5,0.5)$ with increasing
energy, is consistent with the scattering from AFM spin waves.
According to Eq.\ \ref{eq:2}, the scattering from spin wave
excitations of the bilayer Cu spins is expected to be virtually
isotropic in the plane (i.e.\ a ring of scattering) because at
these energies the spin wave wavevector extends only about halfway
to the Brillouin zone boundary. The experimental signal is
consistent with a ring centred on $(0.5,0.5)$ once the statistical
scatter of the data is taken into consideration.

Under the conditions of the measurement the out-of-plane
wavevector component at $E = 95$\,meV is $Q_z = 5.3$ (in units of
$2\pi/c$), and at $E = 145$\,meV it is $Q_z = 6.6$. Maxima in the
bilayer structure factors for acoustic and optic magnons are found
at $Q_z = 5.1$ and 6.8, respectively, so the scattering is mainly
from acoustic modes in Fig. \ref{fig:2}(a) and from optic modes in
Fig. \ref{fig:2}(b).

\begin{figure}
\begin{center}
\includegraphics[width=7.5cm,bbllx=112,bblly=213,bburx=490,
bbury=629,angle=0,clip=]{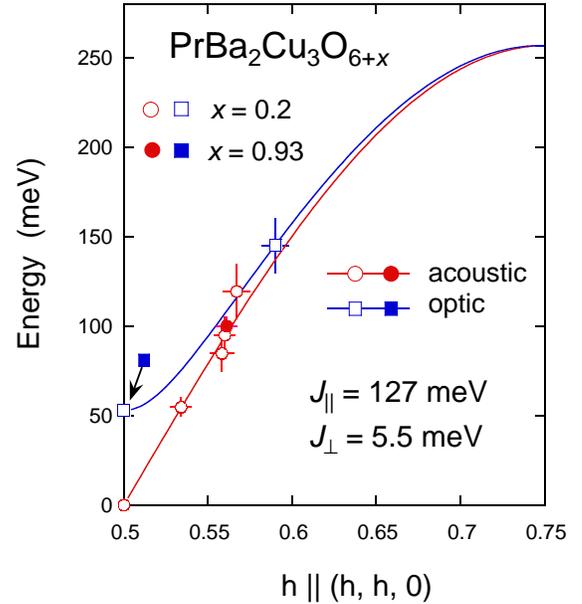} \caption{(Color online).
In-plane magnon dispersion of PrBa$_{2}$Cu$_{3}$O$_{6+x}$. Circles
and squares denote measurements of the acoustic and optic
branches, respectively. Open and closed symbols are for $x=0.2$
and $x=0.93$, respectively. The data points for the optic mode gap
at $h=0.5$ for $x=0.2$ and $x=0.93$ are virtually coincident, as
indicated. Vertical `error bars' indicate the energy range over
which the data were averaged, whereas horizontal error bars show
the experimental uncertainty in the radius of the dispersion
surface.  The lines are the spin wave dispersion relations for a
bilayer antiferromagnet calculated from Eqs.\ \ref{eq:2} and
\ref{eq:3} with $J_{\parallel}=127$\,meV and $J_{\perp} =
5.5$\,meV. \label{fig:4} }
\end{center}
\end{figure}

Figs. \ref{fig:2}(c) and (d) show linear scans extracted from the
intensity maps in Figs. \ref{fig:2}(a) and (b). The scans are made
parallel to $(h,0)$ passing through the AFM zone centre at
$(0.5,0.5)$, and are averaged over a range $\triangle k = 0.08$
(95\,meV) and $\triangle k = 0.1$ (145\,meV) in the $(0,k)$
direction. Each scan contains two peaks, one for each intersection
of the scan with the spin wave dispersion surface which we assume
to be a ring. By fitting a pair of Gaussian functions on a sloping
background to the peaks in these and similar scans in other
directions and from other runs, allowing for the curvature of the
dispersion surface over the averaging width $\triangle k$, we
found the radius of the ring as a function of energy, and hence
arrived at the in-plane spin wave dispersion shown in Fig.\
\ref{fig:4}. We have chosen to plot the dispersion along the
$(h,h,0)$ direction for consistency with published data on
YBCO,\cite{Tranquada-PRB-1989,Reznik-PRB-1996,Hayden-PRB-1996} but
we reiterate that the spin wave dispersion is expected to be
virtually isotropic within the range of wavevectors probed in our
measurement. The points corresponding to mainly acoustic modes and
mainly optic modes are shown with different symbols on Fig.\
\ref{fig:4}, and we have also included the optic mode gap at the
AFM zone centre ($53 \pm 2$\,meV) determined in a previous
measurement on the same PrBCO6.2 crystal.\cite{Lister-PRL-2001}

The magnon dispersion data were compared with the curves for the
acoustic and optic spin wave branches calculated from Eqs.\
\ref{eq:2} and \ref{eq:3}. The two exchange parameters were
adjusted to give the best overall agreement with the experimental
data, and after consideration of the experimental uncertainties we
obtained $J_{\parallel}=127 \pm 10$\,meV and $J_{\perp} = 5.5 \pm
0.9$\,meV. The best-fit dispersion curves are plotted on Fig.\
\ref{fig:4}.

The data collected on oxygenated PrBCO6.93 were more limited than
those just described for PrBCO6.2. This is because the detector
area on HET is considerably smaller than on MAPS, which restricted
the measurement to one Brillouin zone and required two slightly
different settings of the crystal to cover enough of the zone to
include all the signal. Nevertheless, the constant-energy slice
shown in Fig.\ \ref{fig:3}(a) has $Q_z = 5.3$, and so corresponds
to an almost pure acoustic mode. Fig.\ \ref{fig:3}(b) shows a cut
through the antiferromagnetic point $(0.5, 0.5)$ parallel to $(-h,
h)$. This direction was chosen for the cut because the data
recorded in the detector extend furthest along this diagonal, as
can be seen in Fig.\ \ref{fig:3}(b). From the two-Gaussian fit we
obtained the point on the acoustic spin wave dispersion curve
shown on Fig.\ \ref{fig:4}. To within experimental error ($\sim
10$\,\%) the PrBCO6.93 datum lies on the PrBCO6.2 dispersion
curve. In an earlier experiment\cite{Gardiner-ApplPhys-2002} we
determined the optic mode gap for PrBCO6.93 and found it to be $54
\pm 1$\,meV, again consistent with that for PrBCO6.2.

\section{\label{sec:disc}Discussion}

At the start of this work we set out to determine the Cu--Cu
exchange parameters $J_{\parallel}$ and $J_{\perp}$ for PrBCO6+$x$
and to compare them with the corresponding parameters for
YBCO6+$x$. The result $J_{\parallel}=127 \pm 10$\,meV for the
intra-layer exchange parameter of PrBCO6.2 found here is the same
to within experimental error as the values $J_{\parallel}=125 \pm
5$\,meV\cite{Hayden-PRB-1996} and $J_{\parallel}=120 \pm
20$\,meV\cite{Shamoto-PRB-1993} determined by neutron inelastic
scattering for YBCO6.15. There is also good agreement with the
values $J_{\parallel}\approx 115$\,meV for PrBCO6+$x$ ($x \approx
0$), Ref.\ \onlinecite{Yoshida-PRB-1990}, and
$J_{\parallel}\approx 120$\,meV for YBCO6+$x$ ($x \approx 0$),
Ref.\ \onlinecite{Lyons-PRL-1988}, derived from two-magnon Raman
scattering, which provides some support for the models used to
describe the rather broad two-magnon Raman spectra.

On the other hand, the values of the inter-layer exchange
parameter for PrBCO6+$x$ and YBCO6+$x$ are significantly
different. We obtain $J_{\perp} = 5.5 \pm 0.9$\,meV for PrBCO6.2,
compared with $J_{\perp} = 9 - 10$\,meV for YBCO6.2 (Ref.\
\onlinecite{Reznik-PRB-1996}) and $J_{\perp} = 11 \pm 2$\,meV for
YBCO6.15 (Ref.\ \onlinecite{Hayden-PRB-1996}). Our own
measurements on a crystal of YBCO6.2 gave $J_{\perp} = 13 \pm
2$\,meV. One factor that might influence $J_{\perp}$ is the
inter-layer separation, which is about 4\% larger in PrBCO6+$x$
than in YBCO6+$x$ reflecting the larger size of the Pr$^{3+}$ ion
relative to Y$^{3+}$. It would be surprising, however, if this
accounts for a factor 2 difference in $J_{\perp}$ between PrBCO
and YBCO.

As mentioned in the introduction, it has been proposed that
superconductivity is suppressed in PrBCO because holes become
localized in hybridized Pr--O bonds\cite{FR-LM-PRL}. Naively, one
might expect such a change in the electronic structure to have an
influence on the Cu--Cu exchange interactions. The invariance of
$J_{\parallel}$ to the replacement of Y by Pr implies that the
in-plane AFM superexchange interaction is unaffected by any
changes in the electronic structure associated with Pr 4$f$ -- O
2$p$ hydridization. This does not present any obvious difficulties
since the superexchange is mediated by the O 2$p_{\sigma}$
orbitals which lie in the layers, whereas the proposed
hybridization scheme\cite{FR-LM-PRL} involves O 2$p_{\pi}$
orbitals which are oriented perpendicular to the layers. Following
the same reasoning, it is tempting to connect the large difference
in $J_{\perp}$ between PrBCO and YBCO as evidence for
hybridization-induced changes in the electronic structure within
the bilayer. However, if the $J_{\perp}$ exchange were mediated by
hybridized Pr--O bonds that also accommodate doped holes then it
is difficult to explain the apparent insensitivity of $J_{\perp}$
to doping. A fuller analysis of these observations will require a
proper understanding of the mechanism for the inter-layer exchange
in the bilayer cuprates. What we can say, though, is that the
substitution of Y by Pr does have a very significant influence on
the inter-layer coupling.

Before concluding it is worth mentioning that although we have
found no difference between the spin wave spectrum of PrBCO6.2 and
PrBCO6.93 below 100\,meV, there may still be differences at higher
energies. Indeed, the two-magnon Raman peak, which is sensitive to
short wavevelength spin fluctuations, was found shifted to lower
energies in spectra from PrBCO7 compared with PrBCO6,
\cite{Yoshida-PRB-1990,Rubhausen-PRB-1996} leading to a
significantly smaller value $J_{\parallel} \approx 95$\,meV than
obtained from our neutron scattering measurements. One explanation
for this apparent discrepancy could be that the magnetic
excitations in hole-doped PrBCO may be influenced by magnon--hole
interactions at higher energies. Neutron scattering measurements
of the magnetic excitations up to the one-magnon zone boundary
could therefore be very interesting.

\section{\label{sec:conc}Conclusions}

In this work we have shown that the Cu spin excitation spectrum in
PrBCO is well described up to $\sim$150\,meV by the spin wave
model for a bilayer antiferromagnet. In this energy range the spin
excitations do not change with doping. The in-plane superexchange
parameter $J_{\parallel}$ has the same value in undoped PrBCO as
in undoped YBCO, whereas the inter-layer exchange parameter
$J_{\perp}$ is a factor of 2 smaller in PrBCO than in YBCO. An
understanding of these results could provide useful insight into
the differences between the electronic structure of PrBCO and
YBCO, and hence into the question of why YBCO has a
superconducting ground state whereas PrBCO does not.

\begin{acknowledgments}
Financial support was provided by the Engineering and Physical
Sciences Research Council of Great Britain and the Danish
Technical Research Council under the Framework Programme on
Superconductivity.
\end{acknowledgments}


\begin{references}

\bibitem{Allenspach-Handbook-2001}
P. Allenspach and M. B. Maple, in {\it Handbook on the Physics and
Chemistry of Rare Earths}, edited by K. A. Gschneidner Jr., L.
Eyring, and M. B. Maple, Vol. 31 (North-Holland, Amsterdam,), p.
163.

\bibitem{Magnetic-order-YBCO}
An unusual type of spin density wave state has been reported in
underdoped YBCO by Y. Sidis, C. Ulrich, P. Bourges, C. Bernhard,
C. Niedermayer, L. P. Regnault, N. H. Andersen, and B. Keimer,
Phys. Rev. Lett. {\bf 86}, 4100 (2001); and by H. A. Mook, P. Dai,
and F. Do\u{g}an, Phys. Rev. B {\bf 64}, 012502 (2001).

\bibitem{radousky}
H. B. Radousky, J. Mater. Res. {\bf 7}, 1917 (1992).

\bibitem{Boothroyd-JAlloysCmpds-2000}
A. T. Boothroyd, J. Alloys Compd. {\bf 303--304}, 489 (2000).

\bibitem{Zou-JJAP-1997}
Reports of superconductivity in single crystals of PrBCO6+$x$
grown by the floating-zone method have yet to be confirmed --- see
Z. Zou, K. Oka, T. Ito, and Y. Nishihara, Jpn. J. Appl. Phys. {\bf
36}, L18 (1997).

\bibitem{PrBCO-TN}
I. Felner, U. Yaron, I. Nowik, E. R. Bauminger, Y. Wolfus, E. R.
Yacoby, G. Hilscher, and N. Pillmayr, Phys. Rev. B {\bf 40}, 6739
(1989); D. W. Cooke, R. S. Kwok, M. S. Jahan, R. L. Lichti, T. R.
Adams, C. Boekema, W. K. Dawson, A. Kebede, I. Schwegler, J. E.
Crow, and T. Mihalisin, J. Appl. Phys. {\bf 67}, 5061 (1990).

\bibitem{Kebede_PRB-1989}
A. Kebede, C. S. Jee, J. Schwegler, J. E. Crow, T. Mihalisin, G.
H. Myer, R. E. Salomon, P. Schlottmann, M. V. Kuric, S. H. Bloom,
and R. P. Guertin, Phys. Rev. B {\bf 40}, 4453 (1989); A. Kebede,
J. P. Rodriquez, I. Perez, T. Mihalisin, G. Myer, J. E. Crow, P.
P. Wise, and P. Schlottmann, J. Appl. Phys. {\bf 69}, 5376 (1992).

\bibitem{Li-PRB-1989}
W.-H. Li, J. W. Lynn, S. Skanthakumar, T. W. Clinton, A. Kebede,
C. S. Jee, J. E. Crow, and T. Mihalisin, Phys. Rev. B {\bf 40},
R5300 (1989).

\bibitem{Wortmann-Felner-1990}
G. Wortmann and I. Felner, Solid State Commun. {\bf 75}, 981
(1990).

\bibitem{Boothroyd-Cu-Pr-coupling}
A. T. Boothroyd, A. Longmore, N. H. Andersen, E. Brecht, and Th.
Wolf, Phys. Rev. Lett. {\bf 78}, 130 (1997); J. P. Hill, A. T.
Boothroyd, N. H. Andersen, E. Brecht, and Th. Wolf, Phys. Rev. B
{\bf 58}, 11211 (1998); J. P. Hill, D. F. McMorrow, A. T.
Boothroyd, A. Stunault, C. Vettier, L. E. Berman, M. v.
Zimmermann, and Th. Wolf, Phys. Rev. B {\bf 61}, 1251 (2000).

\bibitem{FR-LM-PRL}
R. Fehrenbacher and T. M. Rice, Phys. Rev. Lett. {\bf 70}, 3471
(1993); A. I. Liechtenstein and I. I. Mazin, Phys. Rev. Lett. {\bf
74}, 1000 (1995).

\bibitem{Boothroyd-PhysicaC-1993}
A. T. Boothroyd, S. M. Doyle, and R. Osborn, Physica C{\bf 217},
425 (1993).

\bibitem{Lister-PRL-2001}
S. J. S. Lister, A. T. Boothroyd, N. H. Andersen, B. H. Larsen, A.
A. Zhokhov, A. N. Christensen, and A. R. Wildes, Phys. Rev. Lett
{\bf 86}, 5994 (2001).

\bibitem{Gardiner-ApplPhys-2002}
C. H. Gardiner, S. J. S. Lister, A. T. Boothroyd, N. H. Andersen,
A. A. Zhokhov, A. Stunault, and A. Hiess, Appl. Phys. A {\bf 74}
[Suppl.], S898 (2002).

\bibitem{CEF-excitations}
G. L. Goodman, C.-K. Loong, and L. Soderholm, J. Phys.: Condens.
Matter {\bf 3}, 49 (1991); H.-D. Jostarndt, U. Walter, J.
Harnischmacher, J. Kalenborn, A. Severing, and E. Holland-Moritz,
Phys. Rev. B {\bfseries 46}, 14872 (1992); G. Hilscher, E.
Holland-Moritz, T. Holubar, H.-D. Jostarndt, V. Nekvasil, G.
Schaudy, U. Walter, and G. Fillion, Phys. Rev. B {\bf 49}, 535
(1994).

\bibitem{Gardiner-PRB-2004}
C. H. Gardiner, A. T. Boothroyd, B. H. Larsen, W. Reichardt, A. A.
Zhokhov, N. H. Andersen, S. J. S. Lister, and A. R. Wildes, Phys.
Rev. B {\bf 69}, 092302 (2004).

\bibitem{annealing}
E. D. Specht, C. J. Sparks, A. G. Dhere, J Brynestad, O. B. Cavin,
D. M. Kroeger, H. A. Oye, Phys. Rev. B {\bf 37}, 7426 (1988); N.
H. Andersen, B. Lebech, and H. F. Poulsen Physica C {\bf 172}, 31
(1990); P. Schleger, W. N. Hardy, B. X. Yang, Physica C {\bf 176},
261 (1991).

\bibitem{Parfionov-Konovalov}
O. E. Parfionov and A. A. Konovalov, Physica C {\bf 202}, 385
(1992).



\bibitem{Longmore-PRB-1996}
A. Longmore, A. T. Boothroyd, C. Changkang, H. Yongle, M. P.
Nutley, N. H. Andersen, H. Casalta, P. Schleger, and A. N.
Christensen, Phys. Rev. B {\bf 53}, 9382 (1996).

\bibitem{Uma-JPCM-1998}
S. Uma, W. Schnelle, E. Gmelin, G. Rangarajan, S. Skanthakumar, J.
W. Lynn, R. Walter, T. Lorenz, B. Bfichner, E. Walker, and A. Erb,
J. Phys.: Condens. Matter {\bf 10}, L33 (1998).

\bibitem{Windsor}
C. G. Windsor, {\it Pulsed Neutron Scattering} (Taylor \& Francis,
London, U.K., 1981), p327.

\bibitem{Squires}
G. L. Squires, {\it Introduction to the Theory of Thermal Neutron
Scattering} (Cambridge University Press, Cambridge, U.K., 1978).

\bibitem{Tranquada-PRB-1989}
J. M. Tranquada, G. Shirane, B. Keimer, S. Shamoto, and M. Sato,
Phys. Rev. B {\bf 40}, 4503 (1989).

\bibitem{Reznik-PRB-1996}
D. Reznik, P. Bourges, H. F. Fong, L. P. Regnault, J. Bossy, C.
Vettier, D. L. Milius, I. A. Aksay, and B. Keimer, Phys. Rev. B
{\bf 53}, R14741 (1996).

\bibitem{Hayden-PRB-1996}
S. M. Hayden, G. Aeppli, T. G. Perring, H. A. Mook, F. Do\u{g}an,
Phys. Rev. B {\bf 54}, R6905 (1996).

\bibitem{Shamoto-PRB-1993}
S. Shamoto, M. Sato, J. M. Tranquada, B. J. Sternlieb, and G.
Shirane, Phys. Rev. B {\bf 48}, 13817 (1993).

\bibitem{Yoshida-PRB-1990}
M. Yoshida, N. Koshizuka, and S. Tanaka, Phys. Rev. B {\bf 42},
R8760 (1990).

\bibitem{Lyons-PRL-1988}
K. B. Lyons, P. A. Fleury, L. F. Schneemeyer, and J. V. Waszczak,
Phys. Rev. Lett. {\bf 60}, 732 (1988).

\bibitem{Rubhausen-PRB-1996}
M. R\"{u}bhausen, N. Dieckmann, A. Bock, U. Merkt, W. Widder, and
H. F. Braun, Phys. Rev. B {\bf 53}, 8619 (1996).



\end{references}
\end{document}